\newcommand{\gsim}{\lower.7ex\hbox{$
    \;\stackrel{\textstyle>}{\sim}\;$}}
\newcommand{\lsim}{\lower.7ex\hbox{$
    \;\stackrel{\textstyle<}{\sim}\;$}}
\def\order#1{{\cal O}\left(#1\right)}
\def\eq#1{(\ref{#1})}

\newcommand{\ba}{\begin{eqnarray}}
\newcommand{\ea}{\end{eqnarray}}
\def\dd{{\rm d}}

\documentclass[12pt]{article}
\usepackage{epsfig}

\newcommand\pubnumber{Alberta Thy 11-00\\ BNL-HET-00/35}
\newcommand\pubdate{October 2000}
\newcommand\hepnumber{hep-ph/0010194}

\def\Title#1{\begin{center} {\Large\bf #1 } \end{center}}
\def\Author#1{\begin{center}{ \sc #1} \end{center}}
\def\Address#1{\begin{center}{ \it #1} \end{center}}
\def\andauth{\begin{center}{and} \end{center}}

\newcommand\pubblock{\rightline{\begin{tabular}{l} \pubnumber\\
         \pubdate\\ \hepnumber \end{tabular}}}
\newenvironment{Abstract}{\begin{quotation}  }{\end{quotation}}
\newenvironment{Presented}{\begin{quotation} \begin{center} 
             Presented at the\end{center}
      \begin{center}\begin{large}}{\end{large}\end{center} \end{quotation}}

\makeatletter
\def\section{\@startsection{section}{0}{\z@}{5.5ex plus .5ex minus
 1.5ex}{2.3ex plus .2ex}{\large\bf}}
\def\subsection{\@startsection{subsection}{1}{\z@}{3.5ex plus .5ex minus
 1.5ex}{1.3ex plus .2ex}{\normalsize\bf}}
\def\subsubsection{\@startsection{subsubsection}{2}{\z@}{-3.5ex plus
-1ex minus  -.2ex}{2.3ex plus .2ex}{\normalsize\sl}}

\renewcommand{\@makecaption}[2]{%
   \vskip 10pt
   \setbox\@tempboxa\hbox{\small #1: #2}
   \ifdim \wd\@tempboxa >\hsize     
       \small #1: #2\par          
     \else                        
       \hbox to\hsize{\hfil\box\@tempboxa\hfil}
   \fi}

 \def\citenum#1{{\def\@cite##1##2{##1}\cite{#1}}}
 
\newcount\@tempcntc
\def\@citex[#1]#2{\if@filesw\immediate\write\@auxout{\string\citation{#2}}\fi
  \@tempcnta\z@\@tempcntb\m@ne\def\@citea{}\@cite{\@for\@citeb:=#2\do
    {\@ifundefined
       {b@\@citeb}{\@citeo\@tempcntb\m@ne\@citea\def\@citea{,}{\bf ?}\@warning
       {Citation `\@citeb' on page \thepage \space undefined}}%
    {\setbox\z@\hbox{\global\@tempcntc0\csname b@\@citeb\endcsname\relax}%
     \ifnum\@tempcntc=\z@ \@citeo\@tempcntb\m@ne
       \@citea\def\@citea{,}\hbox{\csname b@\@citeb\endcsname}%
     \else
      \advance\@tempcntb\@ne
      \ifnum\@tempcntb=\@tempcntc
      \else\advance\@tempcntb\m@ne\@citeo
      \@tempcnta\@tempcntc\@tempcntb\@tempcntc\fi\fi}}\@citeo}{#1}}
\def\@citeo{\ifnum\@tempcnta>\@tempcntb\else\@citea\def\@citea{,}%
  \ifnum\@tempcnta=\@tempcntb\the\@tempcnta\else
  {\advance\@tempcnta\@ne\ifnum\@tempcnta=\@tempcntb \else\def\@citea{--}\fi
    \advance\@tempcnta\m@ne\the\@tempcnta\@citea\the\@tempcntb}\fi\fi}
\makeatother

\begin{document}
\begin{titlepage}
\pubblock

\vfill
\def\thefootnote{\fnsymbol{footnote}}
\Title{The Muon Anomalous Magnetic Moment: \\[5pt]  Standard Model 
Theory and Beyond}
\vfill
\Author{Andrzej Czarnecki}
\Address{ Department of Physics, University of Alberta\\
Edmonton, AB, Canada T6G 2J1\\
 and\\
Physics Department, Brookhaven National Laboratory,\\
Upton, NY 11973, USA}
\andauth
\Author{William J. Marciano}
\Address{ Physics Department, Brookhaven National Laboratory,\\
Upton, NY 11973, USA}

\begin{Abstract}
QED, Hadronic, and Electroweak Standard Model contributions to the
muon anomalous magnetic moment, $a_\mu\equiv (g_\mu-2)/2$, are
reviewed.  Theoretical uncertainties in the prediction $a_\mu^{\rm SM}
= 116\,591\,597(67) \times 10^{-11}$ are scrutinized.  Effects due to
``New Physics'' are described.  Implications of the current
experiment vs.~theory constraint $a_\mu^{\rm exp}-a_\mu^{\rm SM} =
453(465)\times 10^{-11}$ and anticipated near term error reduction to
$\pm 155\times 10^{-11}$ are discussed.
\end{Abstract}
\vfill
\begin{Presented}
5th International Symposium on Radiative Corrections \\ 
(RADCOR--2000) \\[4pt]
Carmel CA, USA, 11--15 September, 2000
\end{Presented}
\vfill
\end{titlepage}
\def\thefootnote{\arabic{footnote}}
\setcounter{footnote}{0}
%


\setlength{\unitlength}{1mm}



\section{Introduction}
\label{sec1}
Leptonic anomalous magnetic moments provide precision tests of the
Standard Model and stringent constraints on potential ``New Physics''
effects.  In the case of the electron, comparing the extraordinary
measurements of $a_e \equiv (g_e-2)/2$ at the University of Washington
\cite{Dehmelt87} 
\ba
a_{e^-}^{\rm exp} &=& 0.001\, 159\, 652 \,   188\, 4(43),
\nonumber \\
a_{e^+}^{\rm exp} &=& 0.001\, 159\, 652 \,   187\, 9(43),
\ea
with the prediction
\cite{Mohr99,Laporta:1997zy,Hughes:1999fp,Czarnecki:1998nd} 
\ba
a_e^{\rm SM} &=& {\alpha\over 2\pi} 
 -0.328\,478\,444\,00 \left( {\alpha\over \pi}\right)^2
 +1.181\,234\,017 \left( {\alpha\over \pi}\right)^3
\nonumber\\
&& -1.5098(384)\left( {\alpha\over \pi}\right)^4 +1.66(3)\times 10^{-12}
\mbox{(hadronic \& electroweak loops)}
\ea
provides the best determination of the fine structure constant
\cite{Kinoshita:1996vz}, 
\ba
\alpha^{-1}(a_e) = 137.035\,999\,58(52).
\label{eq3}
\ea
To test the Standard Model requires an alternative measurement of
$\alpha$ with comparable accuracy.  Unfortunately, the next best
determination of $\alpha$, from the quantum Hall effect \cite{Mohr99},
\ba
\alpha^{-1}(qH) = 137.036\,003\,00(270),
\ea
has a much larger error.   If one assumes that $\left| \Delta a_e^{\rm
New\ Physics}\right| \simeq m_e^2 / \Lambda^2$, where $\Lambda$ is the
scale of ``New Physics'', then the agreement between
$\alpha^{-1}(a_e)$ and $\alpha^{-1}(qH)$ currently probes $\Lambda
\lsim \order{\mbox{100 GeV}}$.  
To access the much more interesting  $\Lambda
\sim \order{\mbox{TeV}}$ region would require an order of magnitude
improvement in $a_{e}^{\rm exp}$ (technically feasible \cite{Gab94}),
an improved calculation of the 4-loop QED contribution to $a_e^{\rm
SM}$ and a much better independent measurement of $\alpha^{-1}$ by
almost two orders of magnitude.  The last requirement, although
difficult, is perhaps most
likely to come \cite{Kinoshita:1996vz}  from combining the already
precisely measured Rydberg constant with  a much better determination
of $m_e$.  

We should note that for ``New Physics'' effects that are linear in the
electron mass, $\Delta a_e^{\rm NP} \sim m_e/\Lambda$, naively, one is
currently probing a much more impressive $\Lambda \sim
\order{10^7\mbox{ GeV}}$ and the possible advances described above
would explore $\order{10^9\mbox{ GeV}}$! However, we subsequently
argue that such linear ``New Physics'' effects are misleading or
unlikely.

The measurement of the muon's anomalous magnetic moment has also been
impressive.  A series of experiments at CERN that ended in 1977 found
\cite{PDG98} 
\ba
a_\mu^{\rm exp} = 116 \, 592 \, 300 (840) \times 10^{-11} \qquad
\mbox{(CERN 1977)}.
\label{eq5}
\ea
More recently, an ongoing experiment (E821) at Brookhaven National
Laboratory has been running with much higher statistics and a very
stable, well measured
magnetic field in its storage ring.  Based on data taken
through  1998, combined with the earlier CERN result in (\ref{eq5}),
it found \cite{Brown:2000sj}
\ba
a_\mu^{\rm exp} = 116 \, 592 \, 050 (460) \times 10^{-11} \qquad
\mbox{(CERN'77+BNL'98)}.
\label{eq6}
\ea
Ongoing analysis of E821's 1999 data is expected to reduce the error
in (\ref{eq6}) to about $\pm 140\times 10^{-11}$ before the end of
this year (2000).  The ultimate goal of the experiment is $\pm
40\times 10^{-11}$, about a factor of 20 improvement relative to the
classic CERN experiments.

Although $a_\mu^{\rm exp}$ is currently about 1000 times less precise
than $a_e^{\rm exp}$, it is much more sensitive to hadronic and
electroweak quantum loops as well as ``New Physics'' effects, since
such contributions \cite{km90} are generally proportional to $m_l^2$.
The $m_\mu^2/m_e^2 \simeq 40\,000$ enhancement more than compensates
for the reduced experimental precision and makes $a_\mu^{\rm exp}$ a
more sensitive probe of short-distance phenomena.  Indeed, as we later
illustrate, a deviation in $a_\mu^{\rm exp}$ from the Standard Model
prediction, $a_\mu^{\rm SM}$, could quite naturally be interpreted as
the appearance of ``New Physics'' such as supersymmetry, an exciting
prospect.  Of course, before making such an interpretation, one must
have a reliable prediction for $a_\mu^{\rm SM}$, an issue that we
address in the next section.

Before leaving the comparison between $a_e^{\rm exp}$ and $a_\mu^{\rm
exp}$, we should remark that for cases where ``New Physics''
contributions to $a_l$ scale as $m_l/\Lambda$, roughly equal
sensitivity in $\Lambda$ ($\sim 10^{7} \mbox{ GeV}$) currently exists
for both types of measurements.  However, as previously mentioned,
such examples are in our view artificial.

\section{Standard Model Prediction For $a_\mu$}
\subsection{QED Contribution}
The QED contribution to $a_\mu$ has been computed through 5 loops
\cite{Czarnecki:1998nd,Mohr99} 
\ba
a_\mu^{\rm QED} 
 &=& {\alpha\over 2\pi} 
 +0.765\,857\,376(27) \left( {\alpha\over \pi}\right)^2
 +24.050\,508\,98(44) \left( {\alpha\over \pi}\right)^3
\nonumber\\
&& +126.07(41)\left( {\alpha\over \pi}\right)^4 
+930(170)\left({\alpha\over \pi}\right)^5. 
\ea
Growing coefficients in the $\alpha/\pi$ expansion reflect the
presence of large $\ln{m_\mu/m_e}\simeq 5.3$ terms coming from
electron loops.  Employing the value of $\alpha$ from $a_e$ in
eq.~(\ref{eq3}) leads to 
\ba
a_\mu^{\rm QED} = 116\,584\,705.7(2.9)\times 10^{-11}.
\label{eq8}
\ea 
The current uncertainty is well below the $\pm 40\times 10^{-11}$
ultimate experimental error anticipated from E821 and should,
therefore, play no essential role in the confrontation between theory
and experiment.

\subsection{Hadronic Loop Corrections}
Starting at $\order{\alpha^2}$, hadronic loop effects contribute to
$a_\mu$ via vacuum polarization.  A first principles QCD calculation
of that effect does not exist.  Fortunately, it is possible to
evaluate the leading effect via the dispersion integral \cite{GRaf}
\ba
a_\mu^{\rm Had} (\mbox{vac. pol.}) = {1\over 4\pi^3}
\int_{4m_\pi^2}^\infty {\dd s} \, 
 K\left(s\right) \, \sigma^0(s)_{e^+e^- \to {\rm hadrons}},
\label{eq9}
\ea
where 
$\sigma^0(s)_{e^+e^- \to {\rm hadrons}}$ means QED vacuum
polarization and other extraneous radiative corrections have been
subtracted from measured cross sections, and 
\ba
K(s) &=& x^2\left(1-{x^2\over 2}\right)
 + (1+x)^2 \left( 1+{1\over x^2}\right)
 \left[ \ln(1+x) -x+{x^2\over 2}\right]
 +{1+x\over 1-x}\, x^2 \ln x
\nonumber \\
x&=& {1-\sqrt{1-4m_\mu^2/s} \over 1+\sqrt{1-4m_\mu^2/s}}.
\ea
Detailed studies of eq.~\eq{eq9} have been carried out by a number of
authors \cite{Alemany:1997tn,Davier:1998si,Davier:1999xy,Davier:1998iz,%
Jeg95,kinoshita85,light}. Here, we employ an analysis due to Davier
and H\"ocker \cite{Davier:1998si,Davier:1999xy,Davier:1998iz} which
finds
\ba
a_\mu^{\rm Had}(\mbox{vac. pol.}) = 6924(62)\times 10^{-11}.
\label{eq10}
\ea
It used experimental $e^+e^-$ data, hadronic tau decays,
perturbative QCD and sum rules to minimize the uncertainty in that
result. The  contributions coming from various energy regions are
illustrated in Table \ref{tab1}.
\begin{table}[htb]
\caption{Contributions to $a_\mu^{\rm Had}(\mbox{vac. pol.})$ from
different energy regions as found by Davier and H\"ocker
\protect\cite{Davier:1998si,Davier:1999xy,Davier:1998iz}.}
\label{tab1} 
\vspace*{3mm}
\begin{center}
\begin{tabular}{l@{\hspace{10mm}}r}
\hline
\hline
\\[-4mm]
$\sqrt{s}$ (GeV)  & $a_\mu^{\rm Had}(\mbox{vac. pol.})\times 10^{11}$
\\[.8mm]
\hline
$2m_\pi - 1.8$  & $6343\pm 60$ \\
$1.8 - 3.7$  & $338.7\pm 4.6$ \\
$3.7 - 5+\psi(1S,2S)$  & $143.1\pm 5.4$ \\
$5-9.3$  & $68.7\pm 1.1$ \\
$9.3-12$  & $12.1\pm 0.5$ \\
$12-\infty$  & $18.0\pm 0.1$ \\
\hline
\\[-4mm]
Total    & $6924\pm 62$\\
\hline
\hline
\end{tabular}
\end{center}
\end{table}

It is clear from Table \ref{tab1} that the final result and its
uncertainty are dominated by the low energy region.  In fact, the
$\rho(770 \mbox{ MeV})$ resonance provides about 72\% of the total
hadronic contribution to $a_\mu^{\rm Had}(\mbox{vac. pol.})$.

To reduce the uncertainty in the $\rho$ resonance region, Davier and
H\"ocker employed $\Gamma(\tau\to \nu_\tau \pi^- \pi^0)/\Gamma(\tau
\to \nu_\tau \bar \nu_e e^-)$ data to supplement $e^+e^-\to
\pi^+\pi^-$ cross-sections.  In the $I=1$ channel they are related by
isospin.  Currently, tau decay data is experimentally more precise.

An issue in the use of tau decay data is the magnitude of isospin
violating corrections due to QED and the $m_d-m_u$ mass difference.  A
short-distance QED correction \cite{Marciano:1988vm}  of about $-2\%$
was applied to the hadronic tau decay data and the
$m_{\pi^\pm}-m_{\pi^0}$  phase space difference is easy to account
for.  Other isospin violating differences are estimated to be about
$\pm 0.5\%$ and included in the hadronic uncertainty.  

Although the error assigned to the use of tau decay data appears
reasonable, it has been questioned
\cite{EidelmanPriv,Jegerlehner:1999hg}.  More recent preliminary
$e^+e^-\to \pi^+\pi^-$ data from Novosibirsk \cite{EidelmanPriv} seems
to suggest a potential 1.5 sigma difference with corrected hadronic
tau decays which would seem to further reduce $a_\mu^{\rm Had}$.  It
is not clear whether the difference is due to additional isospin
violating corrections to hadronic tau decays or radiative corrections
to $e^+e^-\to $ hadrons data which must be accounted for in any
precise comparison \cite{Marciano:1992pr}.  If that difference is
confirmed by further scrutiny, it could lead to a reduction in
$a_\mu^{\rm Had}(\mbox{vac. pol.})$.  Resolution of this issue is
extremely important.  However, we note that a reduction in $a_\mu^{\rm
Had}$ would further increase the $a_\mu^{\rm exp}-a_\mu^{\rm SM}$
difference given in the abstract which is roughly 1 sigma at present.

Evaluation of the 3-loop hadronic vacuum polarization contribution to
$a_\mu$  has been updated to \cite{Krause:1997rf,kinoshita85} 
\ba
\Delta a_\mu^{\rm Had}(\mbox{vac. pol.}) = -100 (6) \times 10^{-11}.
\label{eq11}
\ea
Light-by-light hadronic diagrams have been evaluated using chiral
perturbation theory.  An average
\cite{Davier:1998si,Davier:1999xy,Davier:1998iz} of two recent studies
\cite{Bijnens:1996xf,Hayakawa:1998rq} gives
\ba
\Delta a_\mu^{\rm Had}(\mbox{light-by-light}) = -85(25)\times
10^{-11}.
\label{eq12}
\ea
Adding the contributions in Eqs.~(\ref{eq10}), (\ref{eq11}), and
(\ref{eq12}) leads to the total hadronic contribution
\ba
a_\mu^{\rm Had} = 6739(67)\times 10^{-11}.
\label{eq13}
\ea
The uncertainty in that result represents the main theoretical error
in $a_\mu^{\rm SM}$.  It would be very valuable to supplement the
above evaluation of $a_\mu^{\rm Had}$ with lattice calculations (for
the light-by-light  contribution) and improved $e^+e^-$ data.  A goal of
$\pm 40 \times 10^{-11}$ or smaller appears to be within reach and is
well matched to the prospectus of experiment E821 at Brookhaven which
aims for a similar level of accuracy.

\subsection{Electroweak corrections}
The one-loop electroweak radiative corrections to $a_\mu$ are
predicted in the Standard Model to be 
\cite{Brodsky:1967mv,Burnett67,Jackiw72,fls72,Bars72,ACM72,Bardeen72}
\begin{eqnarray}
\lefteqn{a_\mu^{\rm EW}(\rm 1\,loop) =
{5\over 3}{G_\mu m_\mu^2\over 8\sqrt{2}\pi^2}}
\nonumber\\ && \times
\left[1+{1\over 5}(1-4\sin^2\theta_W)^2
+ {\cal O}\left({m_\mu^2 \over M^2}\right) \right]
\nonumber \\
&& \approx 195 \times 10^{-11}
\label{eq14}
\end{eqnarray}
where $G_\mu = 1.16637(1) \times 10^{-5}$ GeV$^{-2}$,
$\sin^2\theta_W\equiv 1-M_W^2/M_Z^2\simeq 0.223$. 
and $M=M_W$ or
$M_{\rm Higgs}$.  The original goal of E821 at Brookhaven was to
measure that predicted effect at about the 5 sigma level (assuming
further reduction in the hadronic uncertainty).  Subsequently, it was
pointed out \cite{KKSS} that two-loop electroweak contributions are
relatively large due to the presence of $\ln m_Z^2/m_\mu^2\simeq 13.5$
terms.  A full two-loop calculation \cite{CKM96,CKM95}, including
low-energy hadronic electroweak loops  \cite{Peris:1995bb,CKM95}, found for
$m_H \simeq 150$ GeV
\ba
a_\mu^{\rm EW}(\mbox{2 loop}) = -43(4) \times 10^{-11},
\label{eq15}
\ea
where the quoted error is a conservative estimate of hadronic, Higgs,
and higher-order corrections.  Combining eqs.~\eq{eq14} and \eq{eq15}
gives the electroweak contribution
\ba
a_\mu^{\rm EW} = 152(4)\times 10^{-11}.
\label{eq16}
\ea
Higher-order leading logs of the form $(\alpha\ln m_Z^2/m_\mu^2)^n$,
$n=2,3,\ldots$ can be computed via renormalization group techniques
\cite{Degrassi:1998es}.  Due to cancellations, they give a relatively small
$+0.5\times 10^{-11}$ contribution to $a_\mu^{\rm EW}$.  It is safely
included in the uncertainty of eq.~\eq{eq16}.

\subsection{Comparison with Experiment}
The complete Standard Model prediction for $a_\mu$ is 
\ba
a_\mu^{\rm SM}= a_\mu^{\rm QED} + a_\mu^{\rm Had} + a_\mu^{\rm EW}.
\ea
Combining eqs.~\eq{eq8}, \eq{eq13} and \eq{eq16}, one finds
\ba
a_\mu^{\rm SM}= 116\,591\,597(67) \times 10^{-11}.
\ea
Comparing that prediction with the current experimental value in
eq.~\eq{eq6} gives
\ba
a_\mu^{\rm exp}-a_\mu^{\rm SM}= 453\pm 465\times 10^{-11}.
\label{eq19}
\ea
That still leaves considerable room for contributions from ``New
Physics'' beyond the Standard Model.  At (one-sided \cite{onesided}) 
95\% CL, one finds
\ba
-310 \times 10^{-11}\le a_\mu({\rm New\ Physics}) \le 1216 \times 10^{-11}.
\label{eq20}
\ea
That constraint is already significant for theories which give
additional negative contributions to $a_\mu$.  Soon, its range will be
reduced by a factor of 3 when the new E821 result is unveiled.  Will a
clear signal for ``New Physics'' emerge?  As we show in the next section,
realistic examples of ``New Physics'' could quite easily lead to
$a_\mu({\rm New\ Physics})\sim \order{400-500\times 10^{-11}}$ which
would appear as about a 3 sigma effect in the near term and increase
to a 6 or 7 sigma effect as E821 is completed and the hadronic
uncertainties in $a_\mu^{\rm SM}$ are further reduced.

\section{``New Physics'' effects}
In general, ``New Physics'' (i.e.~beyond the Standard Model
expectations) will contribute to $a_\mu$ via quantum loop effects.
Indeed, whenever a new model or Standard Model extension is proposed,
$a_\mu^{\rm exp}-a_\mu^{\rm SM}$ is employed to constrain or rule it
out.  Future improvements in $a_\mu^{\rm exp}$ will make such tests
even more powerful.  Alternatively, they may in fact uncover a
significant deviation indicative of ``New Physics''.  

In this section we describe several generic examples of interesting
``New Physics'' probed by $a_\mu^{\rm exp}-a_\mu^{\rm SM}$. Rather
than attempting to be inclusive, we concentrate on two general
scenarios: 1) Supersymmetric loop effects which can be substantial and
would be heralded as the most likely explanation if a deviation in
$a_\mu^{\rm exp}$ is observed and 2) Models of radiative muon mass
generation which predict $a_\mu({\rm New\ Physics}) \sim m_\mu^2 /
M^2$ where $M$ is the scale of ``New Physics''.  Other examples of
potential ``New Physics'' contributions to $a_\mu$ are only briefly
mentioned.

\subsection{Supersymmetry}

\begin{figure}[thb]
\hspace*{-18mm}
\begin{minipage}{16.cm}
\vspace*{6mm}
\[
\mbox{ 
\begin{tabular}{cc}
\psfig{figure=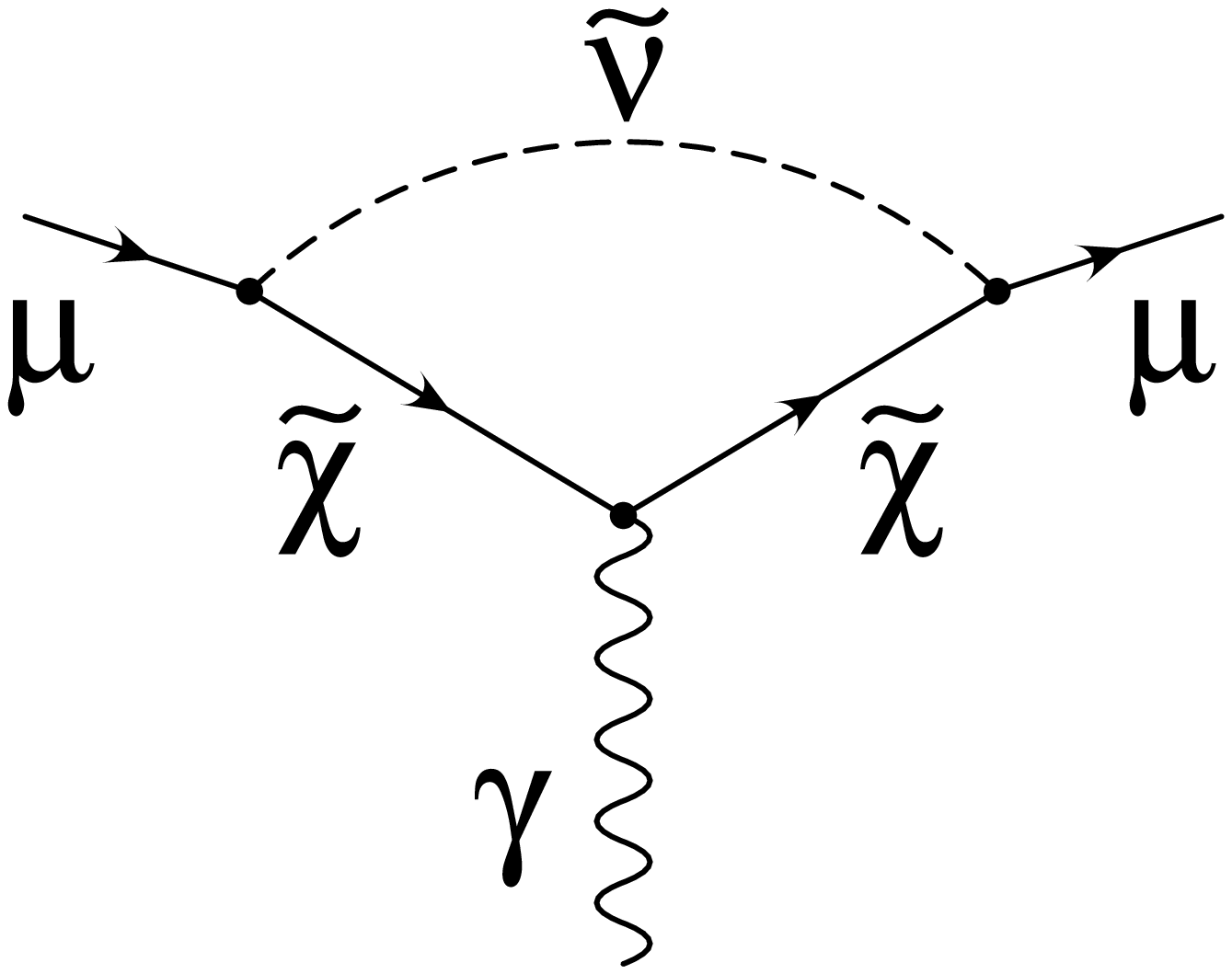,width=35mm,bbllx=72pt,bblly=291pt,%
bburx=544pt,bbury=540pt} 
& \hspace*{10mm}
\psfig{figure=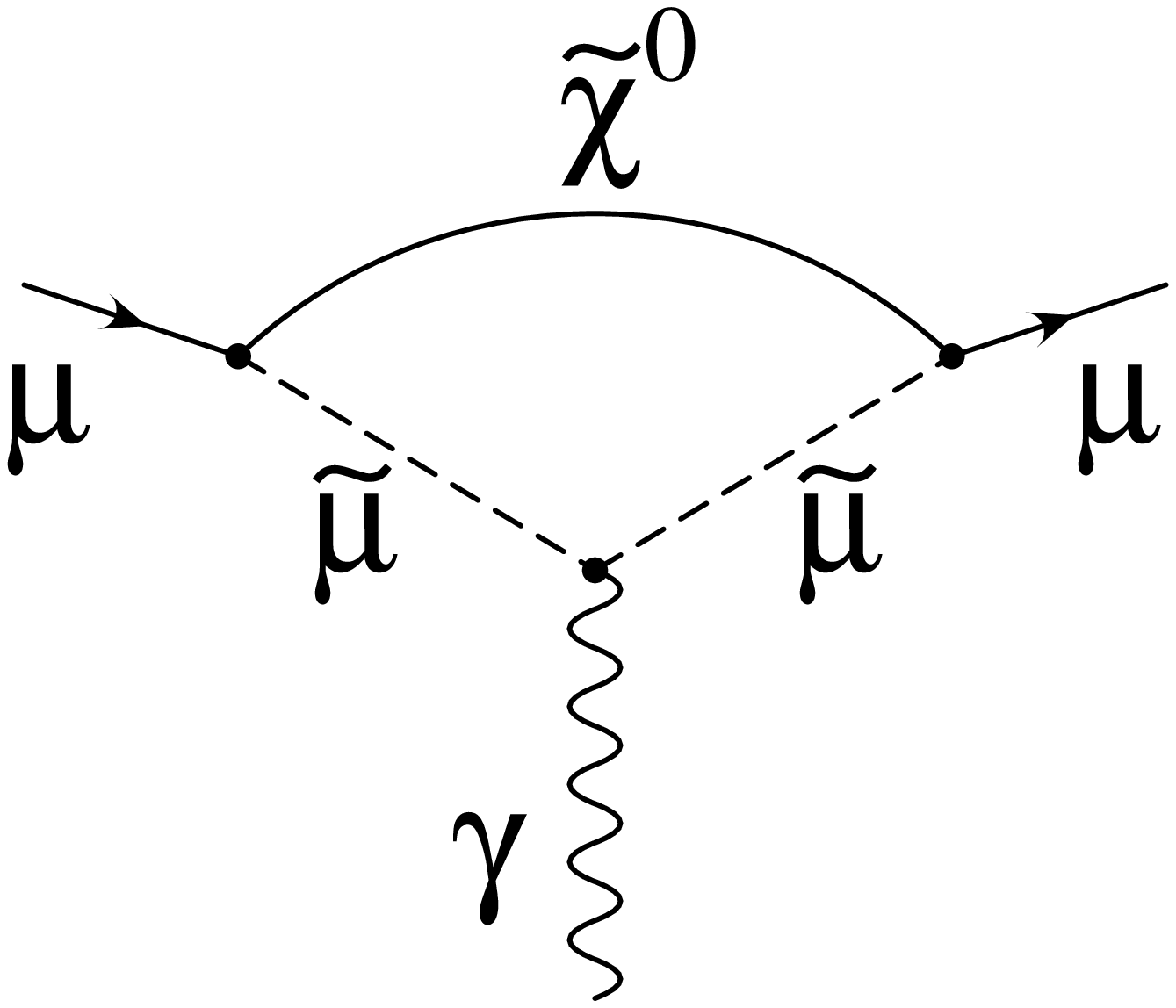,width=35mm,bbllx=72pt,bblly=291pt,%
bburx=544pt,bbury=540pt} 
\\[2mm]
(a) &\hspace*{10mm} (b)
\end{tabular}
}
\]
\end{minipage}
\caption{Supersymmetric loops contributing to the muon anomalous
magnetic moment.}
\label{fig1}
\end{figure}

The supersymmetric contributions to $a_\mu$ stem from
smuon--neutralino and sneutrino-chargino loops (see Fig.~\ref{fig1}).
They include 2 chargino and 4 neutralino states and could in
principle entail slepton mixing and phases. 

Early studies of the supersymmetric contributions $a_\mu^{\rm SUSY}$
were carried out in the context of the minimal SUSY standard model
(MSSM)
\cite{fayet80,Grifols:1982vx,Ellis:1982by,Barbieri:1982aj,Romao:1985pn,%
Kosower:1983yw,Yuan:1984ww,Vendramin:1989rd}, in an $E_6$
string-inspired model \cite{Grifols:1986vr,Morris:1988fm}, and in an
extension of the MSSM with an additional singlet
\cite{Frank:1988yn,Francis:1991pi}.  An important observation was made
in \cite{Lopez:1994vi}, namely that some of the contributions are
enhanced by the ratio of Higgs' vacuum expectation values,
$\tan\beta$, which in some models is large (of order $m_t/m_b\approx
40$).  The main contribution is generally due to the
chargino-sneutrino diagram (Fig.~\ref{fig1}a), which is enhanced by a
Yukawa coupling in the muon-sneutrino-Higgsino vertex.

The leading effect is approximately given in the large $\tan\beta$
limit by 
\ba
\left|a_\mu^{\rm SUSY}\right|
 \simeq {\alpha(M_Z)\over 8\pi\sin^2\theta_W}\,
{m_\mu^2\over \widetilde{m}^2}\,\tan\beta \left( 1-{4\alpha\over
\pi}\ln {\widetilde m\over m_\mu}\right),
\ea
where $\widetilde{m}$ represents a typical SUSY loop mass.  (Chargino-
and sneutrino-masses are assumed degenerate in that expression
\cite{Moroi:1996yh}.)  Also, we have included a 7--8\% suppression
factor due to 2-loop EW effects
\cite{CKM96,Degrassi:1998es}. Numerically, one expects
\ba
\left|a_\mu^{\rm SUSY}\right|
 \simeq 140\times 10^{-11} \left( 100 {\rm\ GeV}\over 
\widetilde{m}\right)^2 \tan\beta,
\label{eq22}
\ea
where $a_\mu^{\rm SUSY}$ generally has the same sign as the
$\mu$-parameter in SUSY models.

Ref.~\cite{Lopez:1994vi} found that E821 will be a stringent test of a 
class of supergravity models.  In the minimal SU(5) SUGRA model,
$\tan\beta$ is severely
constrained by proton decay lifetime and no significant
$a_\mu^{\rm SUSY}$ is possible.  However, extended models, notably
SU(5)$\times$U(1) escape that bound and can induce large effects.  

Supersymmetric effects in $a_\mu$ were subsequently computed in a
variety of models.  Constraints on MSSM were examined in
\cite{Moroi:1996yh,Cho:2000sf}.  MSSM with large CP-violating phases
was studied in \cite{Ibrahim:1999aj}.  Ref.~\cite{Brignole:99}
examined models with a superlight gravitino.  Detailed studies of
$a_\mu^{\rm SUSY}$ were carried out in models constrained by various
assumptions on the SUSY-breaking mechanism: gauge-mediated
\cite{Carena:1997qa,Mahanthappa:1999ta}, 
SUGRA \cite{Nath95,Goto:1999mk,Blazek:1999hb}, and anomaly-mediated
\cite{Chattopadhyay:2000ws}. 

If we simply employ for illustration the large $\tan\beta$ approximate
formula in eq.~\eq{eq22} and the current constraint in eq.~\eq{eq19},
then we find (for positive sgn$(\mu)$)
\ba
\tan\beta \left( {100 \mbox{ GeV}\over \widetilde m}\right)^2 \simeq
3.2 \pm 3.3.
\label{eq23}
\ea
For $\tan\beta \simeq 40$, the non-trivial bound $ \widetilde m \ge
215$ GeV (95\% one-sided CL) follows.  It is anticipated that the
uncertainty in 
that constraint will soon be reduced to $\pm 1$ when the E821 result
is announced.  One can imagine a variety of outcomes and
inferences. If the central value in \eq{eq23} falls to near zero, then
for $\tan\beta\simeq 40$, $ \widetilde m \ge 500$ GeV will result, a
significant constraint.  (Negative sgn$\,\mu$ models are already
tightly constrained.) (Of course, in specific models with
non-degenerate gauginos and sleptons, a more detailed study is
required, but here we only want to illustrate roughly the scale of
supersymmetry probed.)  More interesting would be the case where the
central value in eq.~\eq{eq23} remains fixed and the error is reduced
to $\pm 1$, thereby signaling at a 3 sigma level the presence of ``New
Physics''.  A natural SUSY interpretation would be that sgn$\,\mu$ is
positive, $\tan\beta$ is large $\order{20-40}$ and $ \widetilde m
\simeq 250-350$ GeV or that $\tan\beta$ is moderate $\order{5-10}$ and
$ \widetilde m \simeq 125-180$ GeV.  Either represents a very exciting
prospect with important implications for collider phenomenology as
well as other low energy experiments such as $b\to s \gamma$, $\mu\to
e\gamma$ etc.  Such scenarios are well within the mainstream of SUSY
models. Hence, we anticipate a clear deviation in $a_\mu^{\rm exp}$
from Standard Model expectations to be heralded as strong evidence for
supersymmetry.

\subsection{Radiative Muon Mass Models}
The relatively light masses of the muon and most other known
fundametal fermions suggest that they may be radiatively loop induced
by ``New Physics'' beyond the Standard Model.  Although no compelling
model exists, the concept is very attractive as a natural scenario for
explaining the flavor mass hierarchy.

The basic idea is to start off with a naturally zero bare mass
due to an underlying chiral symmetry.  The symmetry is broken by
quantum loop effects.  They lead to a finite calculable mass which
depends on the mass scales, coupling strengths and dynamics of the
underlying symmetry breaking mechanism.  One generically expects for
the muon 
\ba
m_\mu \propto {g^2\over 16\pi^2} M_F,
\label{eq24}
\ea
where $g$ is some new interaction coupling strength and $M_F\sim
100-1000$ GeV is a heavy scale associated with chiral symmetry
breaking. 

Whatever source of chiral symmetry breaking is responsible for
generating the muon's mass will also give rise to non-Standard Model
contributions in $a_\mu$.  Indeed, fermion masses and anomalous
magnetic moments are intimately connected chiral symmetry breaking
operators.  Remarkably, in such radiative scenarios, the additional
contribution to $a_\mu$ is quite generally given by
\cite{WM:Tennessee,MassMech} 
\ba
a_\mu(\mbox{New Physics}) \simeq C{m_\mu^2 \over M^2}, \qquad C\simeq
\order{1}, 
\label{eq25}
\ea 
where $M$ is some physical high mass scale associated with the ``New
Physics'' and $C$ is a model-dependent number roughly of order 1 (it
can be larger).  $M$ need not be the same scale as $M_F$ in
eq.~\eq{eq24}.  In fact, $M$ is usually a somewhat larger gauge or
scalar boson mass responsible for mediating the chiral symmetry
breaking interaction.  The result in eq.~\eq{eq25} is remarkably
simple in that it is largely independent of coupling strengths,
dynamics, etc.  Furthermore, rather than exhibiting the usual
$g^2/16\pi^2$ loop suppression factor, $a_\mu(\mbox{New Physics})$ is
related to $m_\mu^2/M^2$ by a (model dependent) constant, $C$, roughly
of $\order{1}$.

\begin{figure}[thb]
\hspace*{-5mm}
\begin{minipage}{16.cm}
\vspace*{-2mm}
\[
\raisebox{8mm}{$m_\mu\hspace*{3mm}\simeq$}
\hspace*{5mm}
\psfig{figure=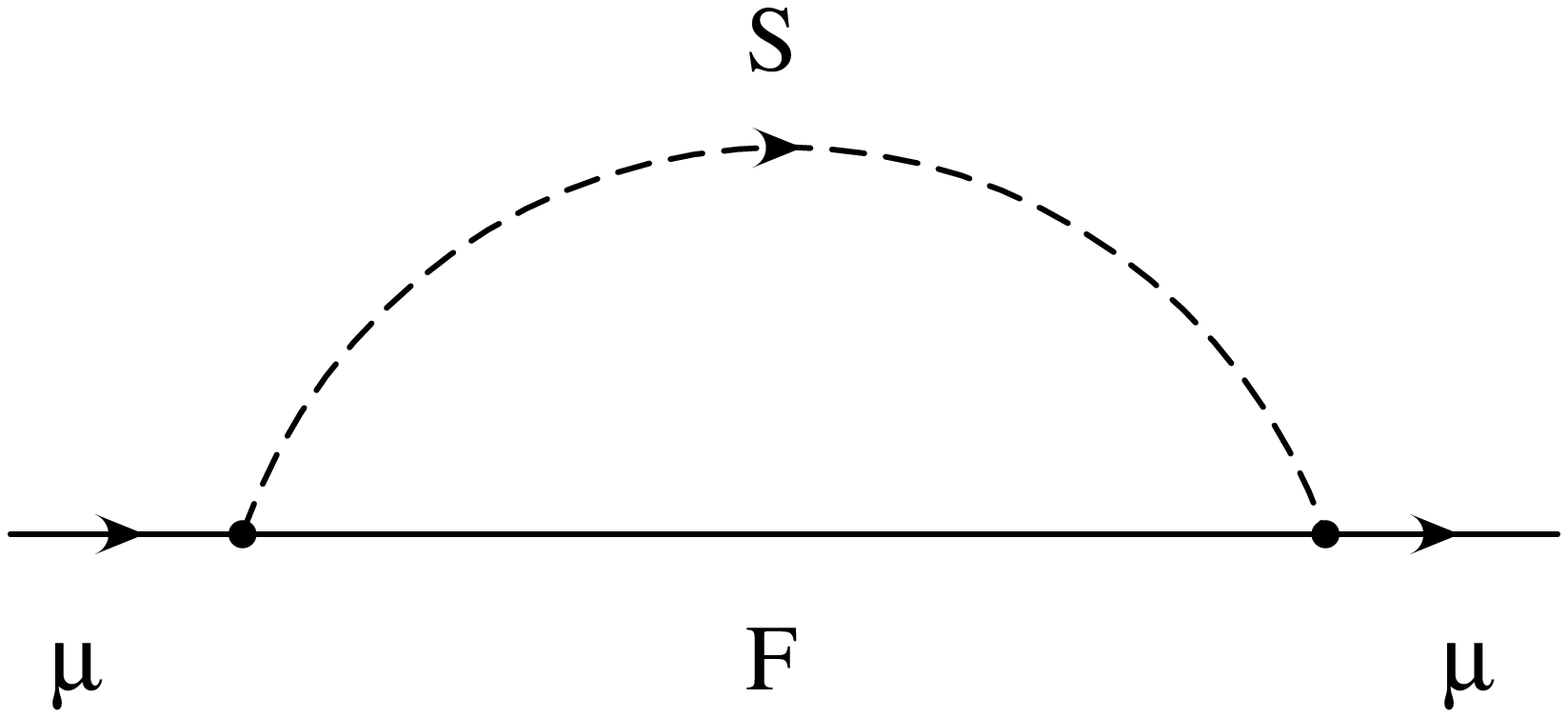,width=55mm}
\hspace*{5mm}
\raisebox{8mm}{+}
\hspace*{5mm}
\psfig{figure=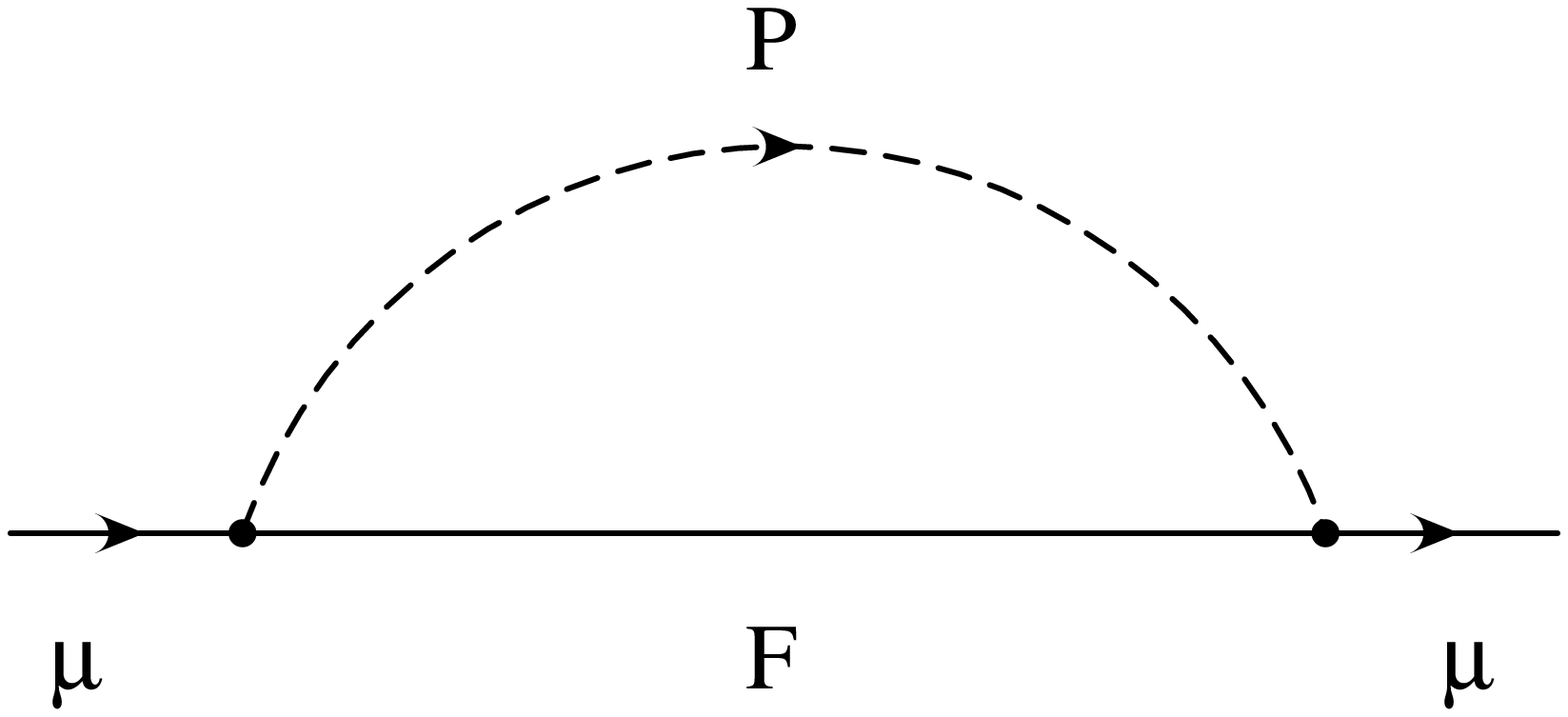,width=55mm}
\]
\end{minipage}
\caption{Example of a pair of one-loop diagrams, which can induce a
finite radiative muon mass.}
\label{fig2}
\end{figure}

To demonstrate how the relationship in eq.~\eq{eq25} arises, we
consider a simple toy model example \cite{MassMech} for muon mass
generation which is graphically depicted in Fig.~\ref{fig2}. 

If the muon is massless in lowest order (i.e.~no bare $m_\mu^0$ is
possible due to a symmetry), but couples to a heavy fermion $F$ via
scalar, $S$, and pseudoscalar, $P$, bosons with couplings $g$ and
$g\gamma_5$ respectively, then the diagrams give rise to
\ba
m_\mu \simeq {g^2\over 16\pi^2} M_F 
             \ln \left( {m_S^2\over m_P^2}\right). 
\label{eq26}
\ea
Note that short-distance ultraviolet divergences have canceled and the
induced mass vanishes in the chirally symmetric limit $m_S=m_P$.  

If we attach a photon to the heavy internal fermion, $F$ (assumed to
have charge $-1$), then a new contribution to $a_\mu$ is also
induced.  For $m_S$, $m_P\gg M_F$, one finds \cite{MassMech}
\ba
a_\mu(\mbox{New Physics}) \simeq {g^2\over 8\pi^2}
 {m_\mu M_F \over m_P^2}
\left( {m_P^2\over m_S^2} \ln {m_S^2\over M_F^2} -  \ln {m_P^2\over
M_F^2}
\right).
\label{eq27}
\ea
It also vanishes in the $m_S=m_P$ chiral symmetry limit.
Interestingly, $a_\mu(\mbox{New Physics})$ exhibits a linear rather
than quadratic dependence on $m_\mu$ at this point.  Recall, that in
section \ref{sec1} we said that such a feature was misleading or
artificial.  Our subsequent discussion should clarify that point. 

Although eqs.~\eq{eq26} and \eq{eq27} both depend on unknown
parameters such as $g$ and $M_F$, those quantities largely cancel when
we combine both expressions.  One finds
\ba
a_\mu(\mbox{New Physics}) & \simeq & C{m_\mu^2 \over m_P^2},
\nonumber \\
C&=& 2\left[ 1-\left(1-{m_P^2 \over m_S^2}\right) \ln {m_S^2 \over M_F^2}/
 \ln {m_S^2 \over m_P^2}
\right],
\ea
where $C$ is very roughly $\order{1}$.  It can actually span a broad
range, depending on the $m_S/m_P$ ratio.  A loop produced
$a_\mu(\mbox{New Physics})$ effect that started out at $\order{
g^2/16\pi^2}$ has been promoted to $\order{1}$ by absorbing the
couplings and $M_F$ factor into $m_\mu$.  Along the way, the linear
dependence on $m_\mu$ has been replaced by a more natural quadratic
dependence. 

A similar relationship, $a_\mu(\mbox{New Physics})\simeq
Cm_\mu^2/M^2$, has been found in more realistic multi-Higgs models
\cite{Babu:1989fg}, dynamical symmetry breaking scenarios such as
extended technicolor \cite{WM:Tennessee,MassMech}, SUSY with soft
masses \cite{Borzumati:1999sp}, etc.  It is also a natural expectation
in composite models
\cite{Brodsky:1980zm,Shaw:1980hk,Gonzalez-Garcia:1996rx} or some
models with large extra dimensions
\cite{Davoudiasl:2000my,Casadio:2000pj}, 
although studies
of
such cases have not necessarily made that same connection.  Basically,
the requirement that $m_\mu$ remain relatively small in the presence
of new chiral symmetry breaking interactions forces $a_\mu(\mbox{New
Physics})$ to effectively exhibit a quadratic $m_\mu^2$ dependence.

For models of the above variety, where $\left| a_\mu(\mbox{New
Physics}) \right| \simeq m_\mu^2/M^2$, the current constraint in
eq.~\eq{eq20} suggests (very roughly) 
\ba
M\gsim \order{1\mbox{ TeV}},
\ea
and that level of sensitivity will expand to about 4 TeV as experiment
E821 improves.  Of course, a non-null finding of $ a_\mu(\mbox{New
Physics}) \simeq 400\times 10^{-11}$ could be interpreted as pointing
to a source of muon mass generation characterized by a mass scale of
$M\sim 1-2$ TeV.  Such a scale of ``New Physics'' could be quite
natural in multi-Higgs models and soft SUSY mass scenarios.  It would
be somewhat low for dynamical symmetry breaking, compositeness and
extra dimension models. 

\subsection{Other ``New Physics'' Examples}
Many other examples of ``New Physics'' contributions to $a_\mu$ have
been considered in the literature.  General analysis in terms of
effective interactions was presented in \cite{Escribano:98}.  Specific
examples include effects due to anomalous $W$ boson magnetic dipole
and electric quadrupole moments
\cite{Herzog:1984nx,Suzuki:1985yh,Grau:1985zh,Beccaria:1998br}, muon
compositeness \cite{Gonzalez-Garcia:1996rx}, extra gauge
\cite{Leveille:1978rc} or Higgs \cite{Krawczyk:1997sm} bosons,
leptoquarks \cite{Couture95,Davidson:1994qk}, bileptons
\cite{Cuypers:1996ia}, 2-loop pseudoscalar effects
\cite{Chang:2000ii}, compact extra dimensions
\cite{Graesser:1999yg,Nath:1999aa} etc.  If a non-Standard Model
effect is uncovered, all will certainly be revisited.

\section{Outlook}
After many years of experimental and theoretical toil, studies of the
muon anomalous magnetic moment are entering a new exciting phase.
Experiment E821 at Brookhaven will soon confront theory at the $\pm
155\times 10^{-11}$ level.  Such sensitivity could start to unveil
``New Physics'' at the several sigma level without too much concern
about theoretical hadronic uncertainties.  Future analysis and runs
would then confirm and refine the discovery.  Theorists would have a
field day.  Alternatively, the experiment could confirm Standard Model
expectations and tighten the bounds on ``New Physics'', a more
traditional role for $a_\mu$.  

Stay tuned, the show is about to begin.

\section*{Acknowledgments}
This work was supported  by the DOE under grant number
DE-AC02-98CH10886.


\begin{thebibliography}{10}
\bibitem{onesided}
We refer to a 95\% one-sided confidence level as roughly equivalent to 90\%
  two-sided.
\end{thebibliography}

\end{document}